\documentclass[referee]{raa}            

\usepackage{graphicx,times}             
\usepackage{natbib}
\usepackage{amssymb,amsmath}
\bibpunct{(}{)}{;}{a}{}{,}

\usepackage[a4paper=true,pagebackref=true,bookmarks=false]{hyperref}
\hypersetup{colorlinks = true, linkcolor = green, anchorcolor = red, citecolor = blue, filecolor = red, pagecolor = red, urlcolor = red}

\begin{document}

   \title{Statistical relations between stellar spectral and luminosity classes
and stellar effective temperature and surface gravity}

   \volnopage{Vol.0 (20xx) No.0, 000--000}      
   \setcounter{page}{1}          

   \author{Oleg Malkov
      \inst{1,2}
   \and Dana Kovaleva
      \inst{2}
   \and Sergey Sichevsky
      \inst{2}
   \and Gang Zhao
      \inst{1}
   }

   \institute{Key Laboratory of Optical Astronomy, National Astronomical Observatories,
             Chinese Academy of Sciences, Beijing 100012, China; {\it malkov@inasan.ru}\\
        \and
             Institute of Astronomy, Russian Academy of Sciences, 48 Pyatnitskaya St., Moscow 119017, Russia\\
\vs\no
   {\small Received~~20xx month day; accepted~~20xx~~month day}}

\abstract{We have determined new statistical relations
to estimate the fundamental
atmospheric parameters of effective temperature and surface gravity,
using MK spectral classification, and vice versa.
The relations were constructed based on the published calibration
tables (for main sequence stars) and observational data from
stellar spectral atlases (for giants and supergiants).
These new relations were applied to field giants
with known atmospheric parameters,
and the results of the comparison of our estimations with
available spectral classification had been quite satisfactory.
}

   \authorrunning{O. Yu. Malkov, D. A. Kovaleva, S. G. Sichevsky \& G. Zhao}  
   \titlerunning{Stellar atmospheric parameters -- spectral class relations}  

   \maketitle

%
%
\section{Introduction}           
\label{sect:intro}

Knowledge of the average effective temperature ($T_{\rm eff}$)
or surface gravity ($\log g$)
of a star of given spectral class and luminosity class is often needed
to solve various problems of astrophysics.
In particular, in the course of our investigation
of interstellar extinction and parameterization of stars,
the necessity was felt to have general statistical relations between the
semi-quantitative parameters spectral class and luminosity class on the one hand, and
$T_{\rm eff}$ and $\log g$, on the other hand. In our study we use
photometry from modern large sky surveys
(DENIS, 2MASS, SDSS, GALEX, UKIDSS, WISE, IPHAS, Pan-STARRS),
containing photometric
(3 to 5 bands) data for $10^7-10^9$ stars. We cross-match
objects in these surveys and can then estimate spectral classes,
luminosity classes, distances ($d$)
and interstellar extinction values ($A_V$),
minimizing the function
\begin{equation}
\chi^2 = \sum_{i=1}^N \left(\frac{m_{obs,i}-m_{calc,i}}{\sigma m_{obs,i}} \right)^2,
\label{equ:functional}
\end{equation}
where $m_{obs,i}$ and $\sigma m_{obs,i}$ are the apparent magnitude
and its observational error, respectively,
in the $i$-th photometric band from a given survey, and
the summation is over up to {\it N} $\sim$ 30 photometric bands, and
\begin{equation}
m_{calc,i} = M_i + 5\log{d} - 5 + A_i.
\label{equ:distance}
\end{equation}
Here $A_i=f(A_V)$ is the extinction in the i-th photometric band,
and can be determined from the interstellar extinction law.
The procedure and the pilot results (including their comparison
with recent Gaia DR2 parallaxes) are described in detail in
\cite{2010MNRAS.401..695M},
\cite{2013AN....334..832S},
\cite{2014AstBu..69..160S},
\cite{2018OAst...27...62M},
\cite{2018Galax...7....7M}.

$M_i=f(\rm{spectral class, luminosity class})$
is the absolute magnitude in i-th photometric band taken
from calibration tables.
To obtain absolute
magnitudes for stars of different spectral classes and luminosity classes
in the corresponding photometric systems $M_i$,
we have used tables of absolute magnitudes in 2MASS, SDSS and GALEX systems
\citet{2007AJ....134.2340K}, \citet{2011AJ....142...23F}.
However, corresponding calibration tables,
which provide stellar absolute magnitudes in a given photometric system
for all spectral classes,
can not be found in the literature for UKIDSS and other surveys.
In the absence of such information,
it is necessary to construct corresponding relations
from theoretical spectral energy distributions (SED) and
photometric system response curves. The best source for
theoretical SEDs are libraries of synthetic spectra
\citep{1997A&AS..125..229L},
\citep{2003IAUS..210P.A20C},
\citep{2008A&A...486..951G},
and the SEDs are computed there for a given set of
atmospheric parameters ($\log T_{\rm eff}$, $\log g$, metallicity)
rather than spectral classes.
Thus, for the decision of this problem, it is necessary to
design relations between spectral class and atmospheric parameters,
for different luminosity classes.
In other words, we have faced questions like, e.g., what are the values of
$T_{\rm eff}$ and $\log g$ for an A2V type star?

Several methods were developed to estimate atmospheric parameters
of effective temperature ($T_{\rm eff}$) and surface gravity ($\log g$),
using photometric indices in the uvby \cite{2011AJ....141..118K},
BVRIJHK \citep{2014Ap&SS.351..229K}, and other colour systems.
On the other hand, several calibration tables, connecting
MK spectral classification with colour indices in
ugriz \citep{2007AJ....134.2398C, 2007AJ....134.2340K},
GALEX FUV/NUV \citep{2010AJ....139.1338F, 2011AJ....142...23F},
JHK \citep{2013ApJS..208....9P},
and other photometric systems was published.
However, there is a lack of relations between MK spectral classification
and atmospheric parameters.
Individual precise measurements of fundamental parameters of $T_{\rm eff}$
and $\log g$ are published in many sources,
however, they are presented mostly for MS-stars (see, e.g., \citep{1995A&A...293..446S}).

The goal of this study is to construct analytical statistical relations
between MK spectral class and the atmospheric parameters ($T_{\rm eff}$, $\log g$)
for principal luminosity classes.
Similar work was done more than thirty years ago by
\cite{1987A&A...177..217D}, where
statistical relations between stellar spectral and luminosity classes
and stellar effective temperature and luminosity (but not surface gravity)
were determined.
Several authors have given interpolation tables presenting
($T_{\rm eff}$, $\log g$) -- (spectral class, luminosity class) relations
(see Section~\ref{sec:data} for references), but we wanted to add more
recent determinations of $T_{\rm eff}$ and $\log g$ to these data.
Also, it appeared necessary to have these relations adapted
to computer-use, which would enhance their usefulness.

The paper is organized as follows.
Observational data and published relations used in the present study
are listed in briefly discussed in Section~\ref{sec:data}.
Construction of (spectral class --- $\log T_{\rm eff}$/$\log g$)
statistical relations for dwarfs, supergiants and giants
is described in Sections~\ref{sec:V}, \ref{sec:I} and~\ref{sec:III},
respectively.
Application of these relations to stars with known metallicity
is discussed in Section~\ref{sec:metallicity}.
Section~\ref{sec:lamost}
contains verification of our results with LAMOST data,
and we draw our conclusions in Section~\ref{sec:conclusions}.

\section{Data sources}
\label{sec:data}

The main source of accurate data on stellar astrophysical parameters
is components of detached double-lined eclipsing binaries. The lists
of such binaries were compiled by
\cite{1980ARA&A..18..115P}, \cite{1988BAICz..39..329H}, \cite{1991A&ARv...3...91A},
\cite{2010A&ARv..18...67T}, \cite{2018MNRAS.479.5491E},
and they contain mostly main sequence stars.

Another useful source of stars with available spectral classification and
known atmospheric parameters is empirical stellar spectral atlases.
In the current study we use data from
ELODIE~\citep{2007astro.ph..3658P},
Indo-US~\citep{2004ApJS..152..251V},
MILES~\citep{2011A&A...532A..95F},
and STELIB~\citep{2003A&A...402..433L}
libraries.

Calibration tables based on observational data were constructed by
\cite{1966ARA&A...4..193J},
\cite{1976asqu.book.....A},
\cite{1980ARA&A..18..115P},
\cite{1992msp..book.....S},
\cite{2013ApJS..208....9P}
(see also Mamajek's data available at
\url{http://www.pas.rochester.edu/~emamajek/EEM_dwarf_UBVIJHK_colors_Teff.txt}
and published in \citep{2012ApJ...746..154P}), and
\cite{2018MNRAS.479.5491E}.
Of these, only \cite{1976asqu.book.....A} and \cite{1992msp..book.....S}
presented $\log g$ values for non-MS stars.

\section{Construction of statistical relations}
\label{sec:relations}

\subsection{Main sequence stars}
\label{sec:V}

To construct analytic (spectral class -- atmospheric parameters) formula
for main sequence stars we have used published relations and
made a polynomial approximation.

\begin{figure}
\centering
\includegraphics[width=7cm]{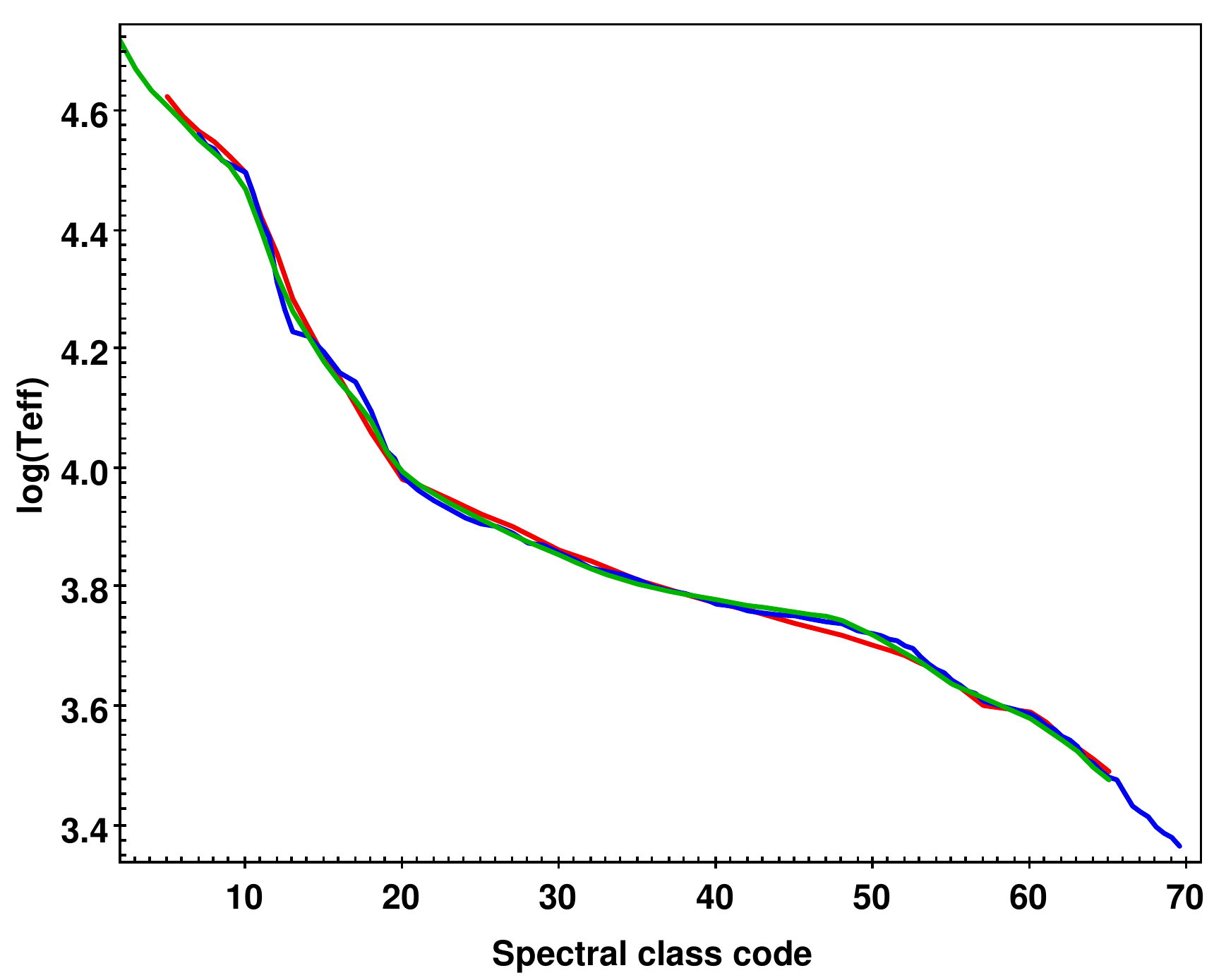}
\includegraphics[width=7cm]{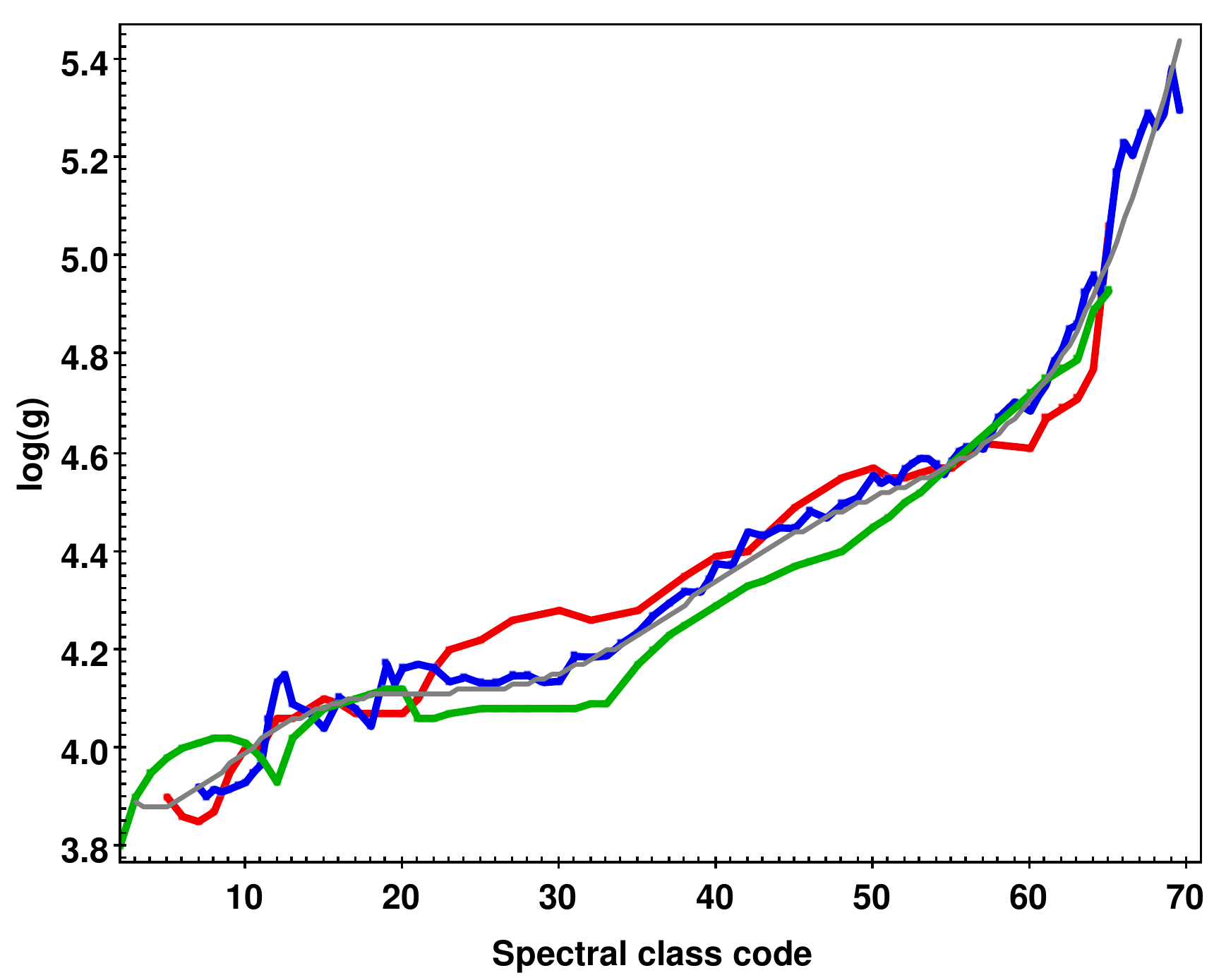}
\caption{Main sequence stars. Spectral class --- effective temperature (left panel)
and spectral class --- surface gravity (right panel) relations.
Spectral class is coded as follows: 3 for O3, ..., 10 for B0, ..., 60 for M0.
\cite{1992msp..book.....S},
\cite{2013ApJS..208....9P}, and
\cite{2018MNRAS.479.5491E}
data are represented by red, blue and green curves, respectively.
Gray curve is the ($\log$ g -- spectral class) relation, approximated by
polynomial (see Table~\ref{tab:main}, Eq.~(T3)).
}
\label{fig:Vtg}
\end{figure}

Spectral class --- effective temperature relations of different authors
demonstrate an excellent agreement (see Fig.~\ref{fig:Vtg}).
To draw an analytical form for the spectral class --- $\log T_{\rm eff}$
relation we have used
\cite{2018MNRAS.479.5491E} data,
with
\cite{2013ApJS..208....9P}
extension to coolest stars.

An agreement of (spectral class --- surface gravity) relations of the same
authors is worse (see Fig.~\ref{fig:Vtg}).
To construct (spectral class --- $\log g$) analytic formula we have used
all three relations.

The results are presented in Table~\ref{tab:main} (Eqs.(T1-T4))
and Fig.~\ref{fig:Vtg}.

\begin{table*}
\centering
\caption{Spectral class ---  effective temperature --- surface gravity relations}
\begin{tabular}{lllllr}
\hline
& & & std.      & valid for & Eq. \\
& & &      dev. &           &     \\

\hline
\multicolumn{4}{l}{\rule{0pt}{12pt}LC=V}\\[2pt]

$\log T_{\rm eff}$ & = &
$      4.80223
   -0.0465961 S
  +0.00157054 S^2    $
&   0.004       & O3--O9 & (T1) \\
$\log T_{\rm eff}$ & = &
$      5.30408
    -0.111312 S
  +0.00284209 S^2
-2.51285e^{-5} S^3     $
&    0.011      & B0--G7 \\
$\log T_{\rm eff}$ & = &
$      3.25745
   +0.0285452 S  
 -0.000388153 S^2    $
&   0.008       & G8--M9 \\

\multicolumn{4}{l}{\rule{0pt}{12pt}}\\[2pt]

S & = &
$     -77.4025
     -208.506 T        
     -72.7616 T^2     $
&     0.36     & $3.38  \le \log T_{\rm eff}  < 3.75  $  & (T2) \\
S & = &
$      13.0566
     +68.6827 T        
     +404.486 T^2      
     +751.011 T^3      
     +497.913 T^4     $
&     0.75     & $3.75  \le \log T_{\rm eff}  < 4.10  $ \\
S & = &
$      5.53554
     -34.2627 T
     -4.78570 T^2
     +191.168 T^3
     +317.065 T^4     $
&     0.34     & $4.10  \le \log T_{\rm eff}  \le 4.72  $ \\

\multicolumn{4}{l}{\rule{0pt}{12pt}}\\[2pt]

$\log g$ & = &
$      4.23248
   +0.0194541 S_1
 +0.000552749 S_1^2
-4.30515e^{-5}  S_1^3 - $ & & \\
         &   &
$-1.09920e^{-6}  S_1^4
+7.61843e^{-8}  S_1^5
+8.20985e^{-10} S_1^6
-3.27874e^{-11} S_1^7  $
&   0.055      & O3--M9.5  & (T3) \\

S & = &
$    -0.117642
      +1.07059 G        
      +192.069 G^2      
      -183.386 G^3      
      +49.7143 G^4     $
&     4.02        & $3.8   \le \log g \le 5.3   $  & (T4) \\

\hline

\multicolumn{4}{l}{\rule{0pt}{12pt}LC=I}\\[2pt]

$\log T_{\rm eff}$ & = &
$       5.37107
      -0.132197 S  
    +0.00447197 S^2
  -7.12416e^{-5} S^3
  +4.17523e^{-7} S^4    $
&     0.049     & O7--M3  & (T5) \\

S & = &
$      5.87386
      -49.0805 T        
      -135.952 T^2      
      -119.090 T^3      
      +124.459 T^4
      +108.708 T^5     $
&     3.14       & $3.45  \le \log T_{\rm eff}  < 4.60  $  & (T6) \\

$\log g$ & = &
$       5.26666
      -0.289286 S  
    +0.00728099 S^2
  -6.33673e^{-5} S^3    $
&      0.485    & O7--M3  & (T7) \\

S & = &
$      5.26199
      -10.2492 G        
      +2.79561 G^2      
     +0.526251 G^3     $ 
&     9.74       & $-0.2   \le \log g \le 3.8   $  & (T8) \\

\hline
\multicolumn{4}{l}{\rule{0pt}{12pt}LC=III}\\[2pt]

$\log T_{\rm eff}$ & = &
$       5.07073
     -0.0757056 S  
    +0.00147089 S^2
  -1.03905e^{-5} S^3    $
&     0.034     & O5--M10  & (T9) \\

S & = &
$     8.49594
     -49.4053 T        
     -191.524 T^2      
     -335.488 T^3      
     -144.781 T^4     $
&     2.59      & $3.45  \le \log T_{\rm eff}  < 4.65  $  & (T10) \\

$\log g$ & = &
$       3.79253
     -0.0136260 S  
   +0.000562512 S^2
  -1.68363e^{-5} S^3    $
&      0.513    & O5--M10  & (T11) \\

S & = &
$      33.3474
      -18.3022 G        
      -5.33024 G^2      
     -0.667234 G^3     $ 
&      7.03      & $-0.5   \le \log g \le 4.7   $  & (T12) \\

\hline
\multicolumn{4}{l}{\footnotesize
S is spectral class code: 3 for O3, ..., 10 for B0, ..., 60 for M0.}\\
\multicolumn{4}{l}{\footnotesize $S_1$ = $S-35$}\\
\multicolumn{4}{l}{\footnotesize T = $\log T_{\rm eff} - 4.6$}\\
\multicolumn{4}{l}{\footnotesize G = $\log g - 3.7$}\\
\end{tabular}
\label{tab:main}
\end{table*}

\subsection{Supergiants}
\label{sec:I}

Supergiants are rare in the Galaxy but due to high intrinsic
luminosity they are relatively numerous in observational catalogues and surveys,
especially at low galactic latitudes. In particular, 276 from 9110 stars
in Bright star catalogue~\citep{1991bsc..book.....H}
are supergiants.

To construct analytic (spectral class -- atmospheric parameters) formula
for supergiant stars, we have used data from 
the empirical stellar spectral atlases
ELODIE~\citep{2007astro.ph..3658P},
Indo-US~\citep{2004ApJS..152..251V},
MILES~\citep{2011A&A...532A..95F},
and STELIB~\citep{2003A&A...402..433L},
and made a polynomial approximation.

Approximating the observational data we do not separate
high (Ia), intermediate (Iab) and low (Ib) luminosity supergiant stars,
and construct overall relations for all supergiants.
Empirical stellar spectral libraries do not provide observational
errors for atmospheric parameters, so
we have assigned equal weight to all stars.

\begin{figure}
\centering
\includegraphics[width=7cm]{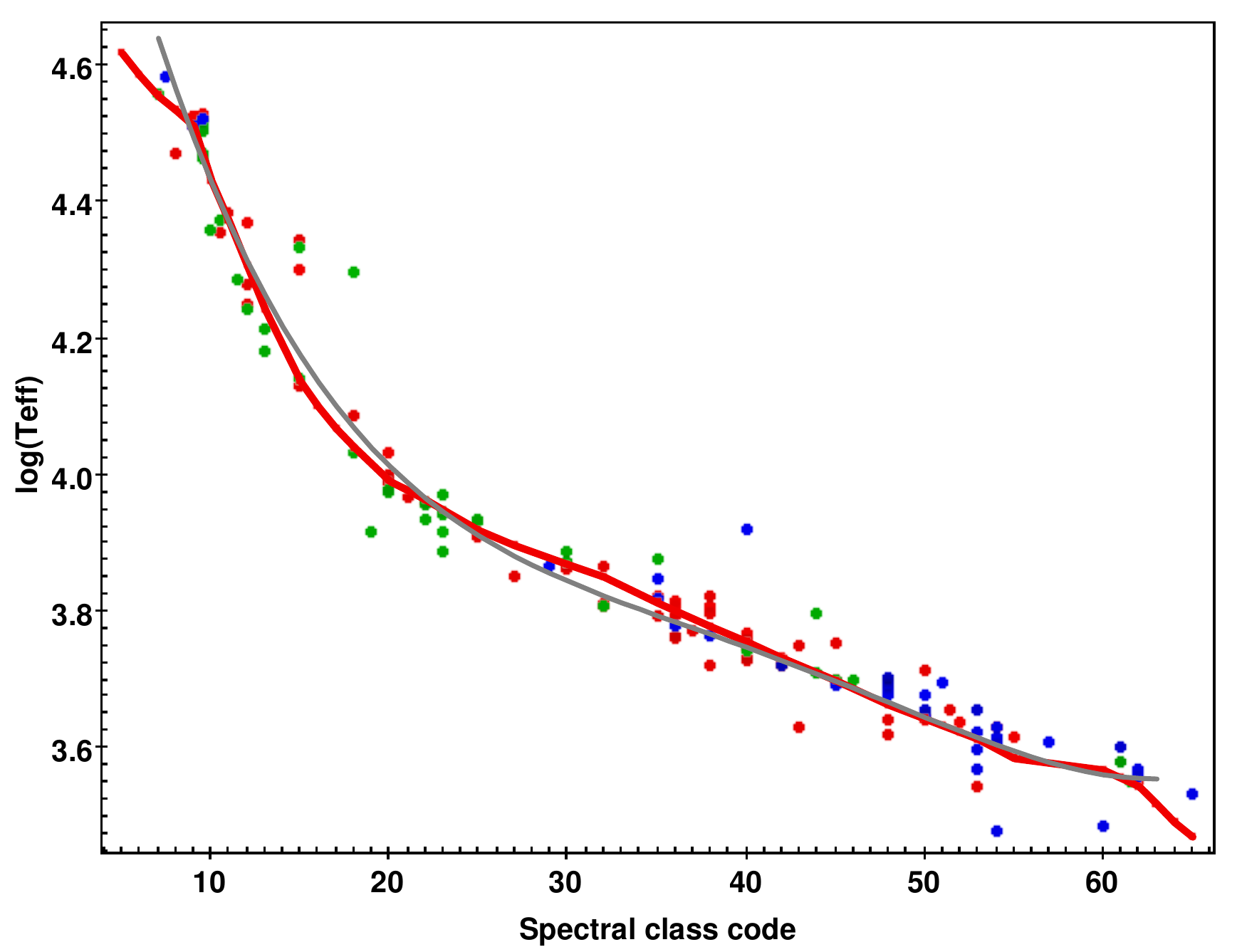}
\includegraphics[width=7cm]{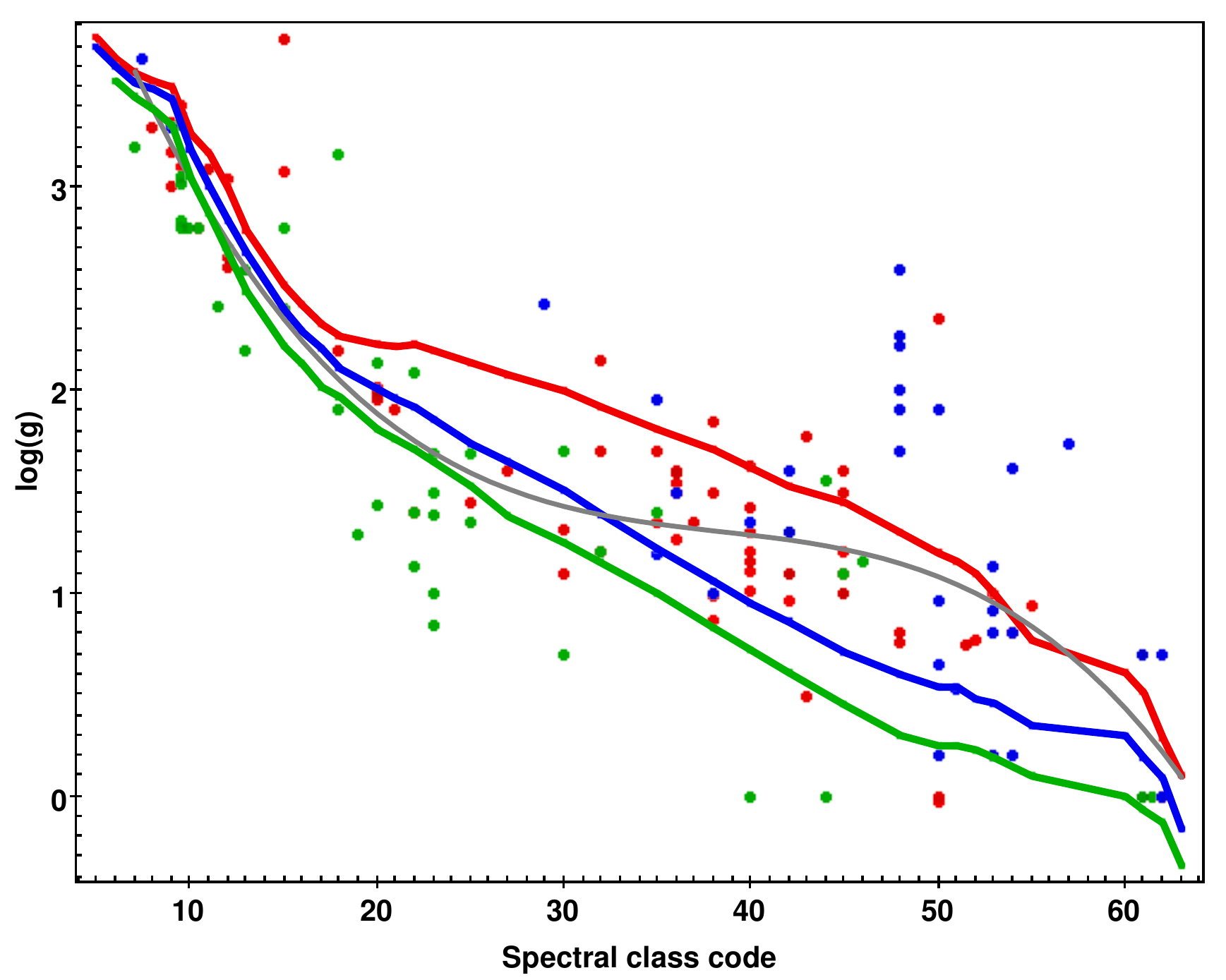}
\caption{Supergiants. Spectral class --- effective temperature (left panel)
and spectral class --- surface gravity (right panel) relations.
Spectral class is coded as follows: 3 for O3, ..., 10 for B0, ..., 60 for M0.
Low (Ib), intermediate (Iab) and high (Ia) luminosity supergiants,
selected from empirical stellar spectral atlases, are represented by
red, blue and green dots, respectively.
Gray curves are the ($\log T_{\rm eff}$ -- spectral class)
and ($\log$ g -- spectral class) relations, approximated by
polynomial (see Table~\ref{tab:main}, Eqs.~(T5) and~(T7), respectively).
Left panel: red curve represents \cite{1992msp..book.....S}
(spectral class -- $\log T_{\rm eff}$) relation for supergiants.
Right panel: green, blue and red curves represent \cite{1992msp..book.....S}
(spectral class -- $\log g$) relations for Ia, Iab and Ib supergiants, respectively.
}
\label{fig:Itg}
\end{figure}

The results are presented in Table~\ref{tab:main} (Eqs.(T5-T8))
and Fig.~\ref{fig:Itg}.

\subsection{Giants}
\label{sec:III}

As in the previous Section,
to construct analytic (spectral class -- atmospheric parameters) formula
for giant stars, we have used data from 
the empirical stellar spectral atlases
ELODIE~\citep{2007astro.ph..3658P},
Indo-US~\citep{2004ApJS..152..251V},
MILES~\citep{2011A&A...532A..95F},
and STELIB~\citep{2003A&A...402..433L}
(besides that a small number of giants from
\cite{2018MNRAS.479.5491E} list was added)
and made a polynomial approximation.

\begin{figure}
\centering
\includegraphics[width=7cm]{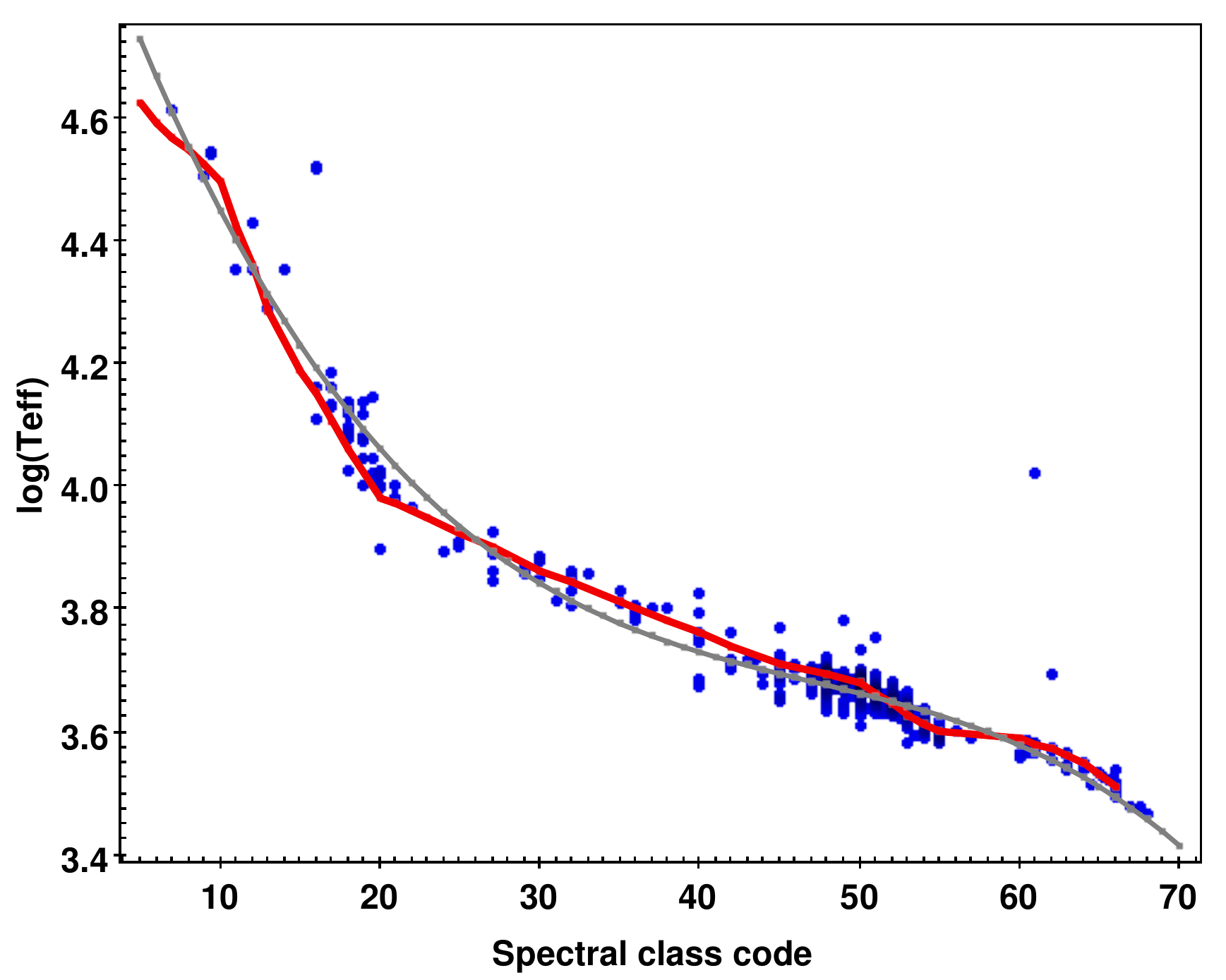}
\includegraphics[width=7cm]{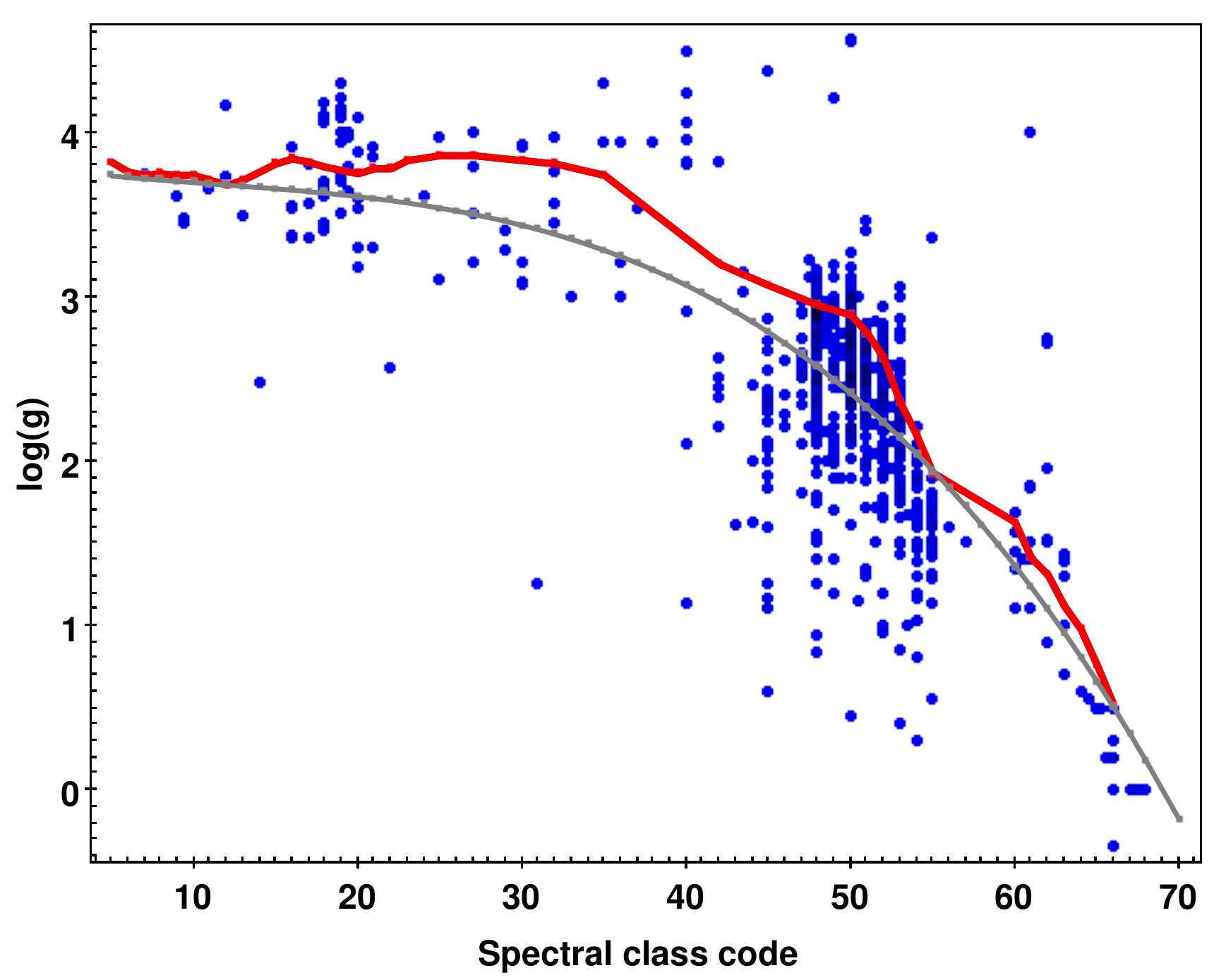}
\caption{Giants. Spectral class --- effective temperature (left panel)
and spectral class --- surface gravity (right panel) relations.
Spectral class is coded as follows: 3 for O3, ..., 10 for B0, ..., 60 for M0.
Dots represent giants, selected from empirical stellar spectral atlases.
Gray curves are the ($\log T_{\rm eff}$ -- spectral class)
and ($\log$ g -- spectral class) relations, approximated by
polynomial (see Table~\ref{tab:main}, Eqs.~(T9) and~(T11), respectively).
Red curves represent \cite{1992msp..book.....S}
(spectral class -- $\log T_{\rm eff}$)
and (spectral class -- $\log g$) relations for giants.
}
\label{fig:IIItg}
\end{figure}

The results are presented in Table~\ref{tab:main} (Eqs.~(T9-T12))
and Figs.~\ref{fig:IIItg}.

\section{Stars with known metallicity}
\label{sec:metallicity}

Spectral classification (in contrast to narrow band photometry)
is a very poor indicator of stellar metallicity [Fe/H].
Also, the number of stars with known chemical abundance is much
less than the number of stars with available spectral
classification and/or $\log T_{\rm eff}$ / $\log g$ parameters.
Last but not least, there is no obvious correlation between
the initial mass (i.e., spectral class) of a star and its metallicity.

Equations in Table~\ref{tab:main} are derived assuming one does not have data
on stellar metallicity.
However, sometimes data on [Fe/H] is available, along with data on $T_{\rm eff}$
and $\log g$. In this case, some corrections can be made to those equations.

\begin{figure}
\centering
\includegraphics[width=8cm]{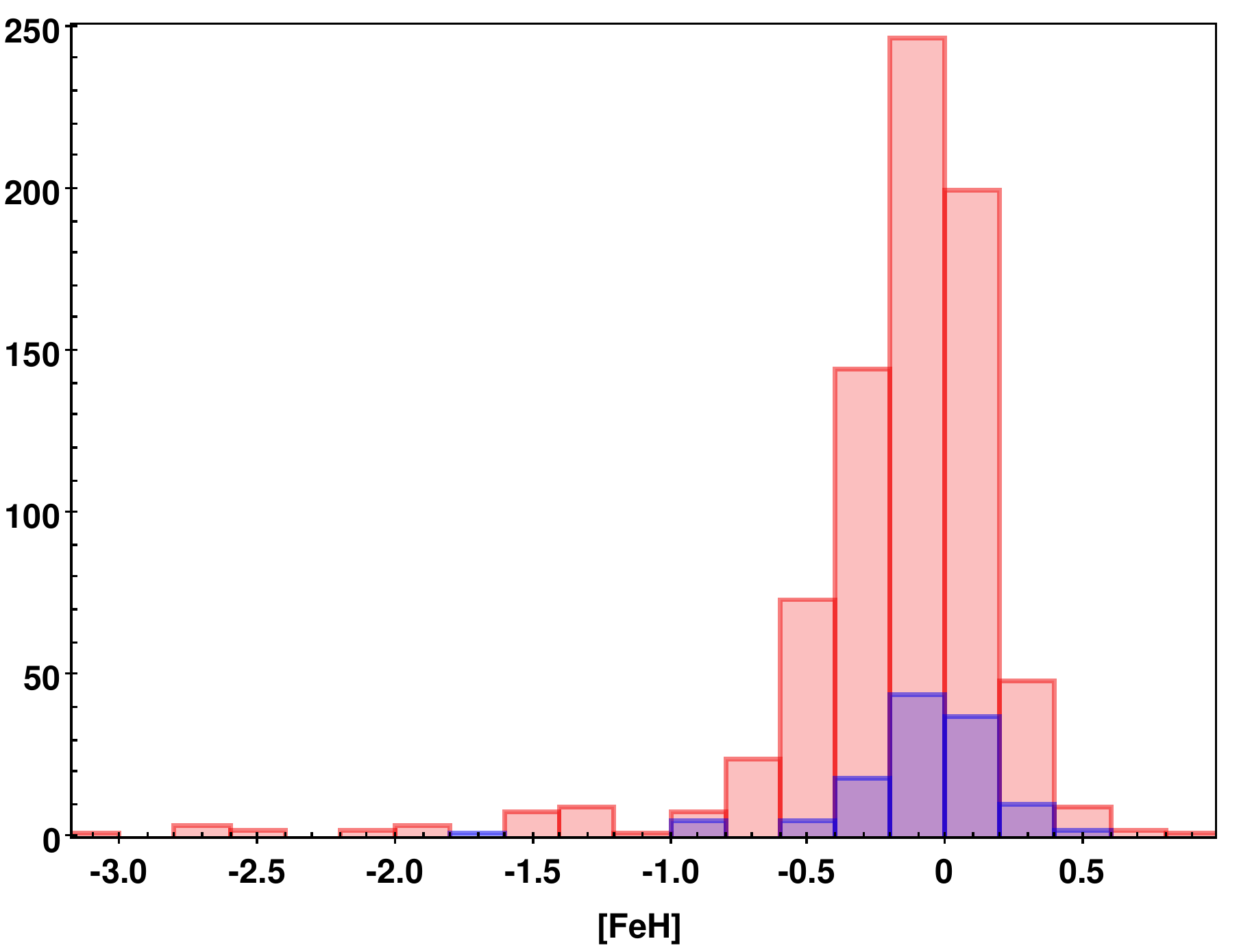}
\caption{Metallicity distribution of 783 giants (red bars) and 122 supergiants (blue bars).
Stars are from ELODIE, MILES, Indo-US, STELIB empirical stellar spectral atlases, and
from \citep{2010A&ARv..18...67T} list.
}
\label{fig:feh}
\end{figure}

\begin{figure}
\centering
\includegraphics[width=7cm]{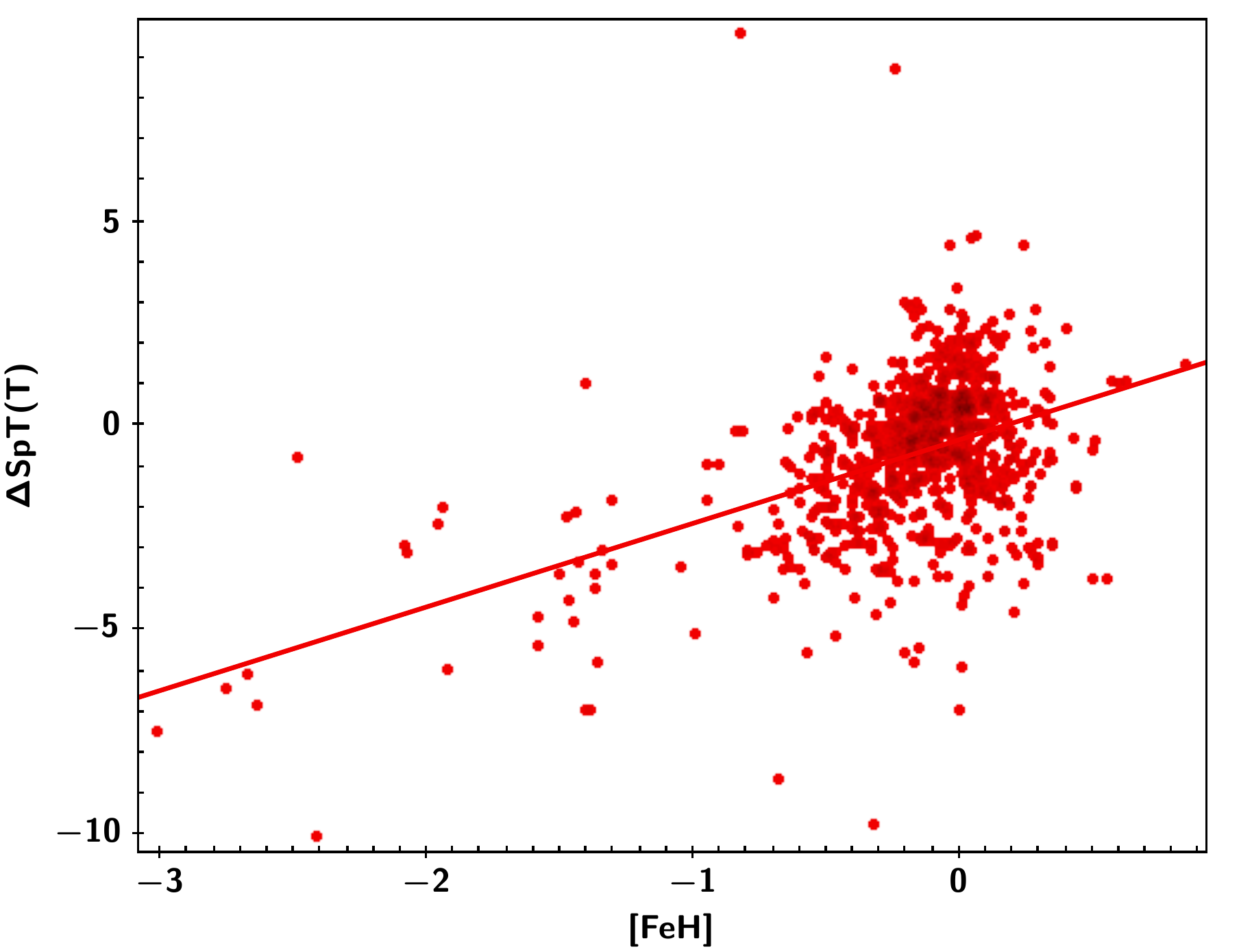}
\includegraphics[width=7cm]{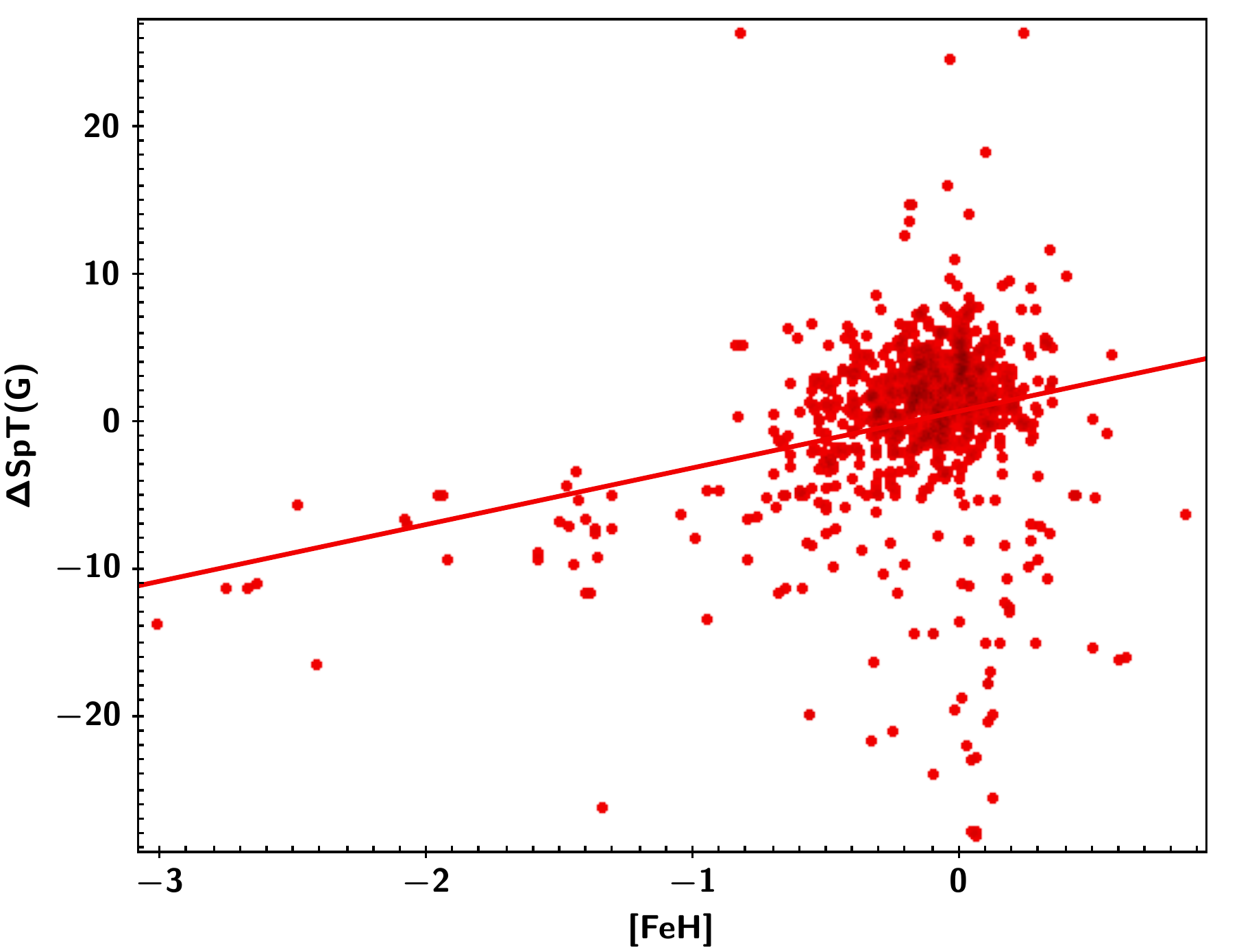}
\caption{Giants. Difference between observed minus computed spectral class
vs. metallicity.
Left panel: $\Delta SpT(T)=SpT_{cat} - SpT(T_{\rm eff})$, where SpT$_{cat}$ is
the spectral class, catalogued in the empirical stellar spectral atlases,
and SpT($T_{\rm eff}$) is the spectral class, calculated from $T_{\rm eff}$ with Eq.~(T10).
Right panel: $\Delta SpT(G)=SpT_{cat} - SpT(\log g)$, where SpT($\log g$)
is the spectral class, calculated from $\log g$ with Eq.~(T12).
Linear fit is shown by the solid line.
The Y-axis is graded so that one unit corresponds to one spectral sub-class
(e.g., the difference between A1 and A2).
}
\label{fig:g-feh-dS}
\end{figure}

\begin{figure}
\centering
\includegraphics[width=7cm]{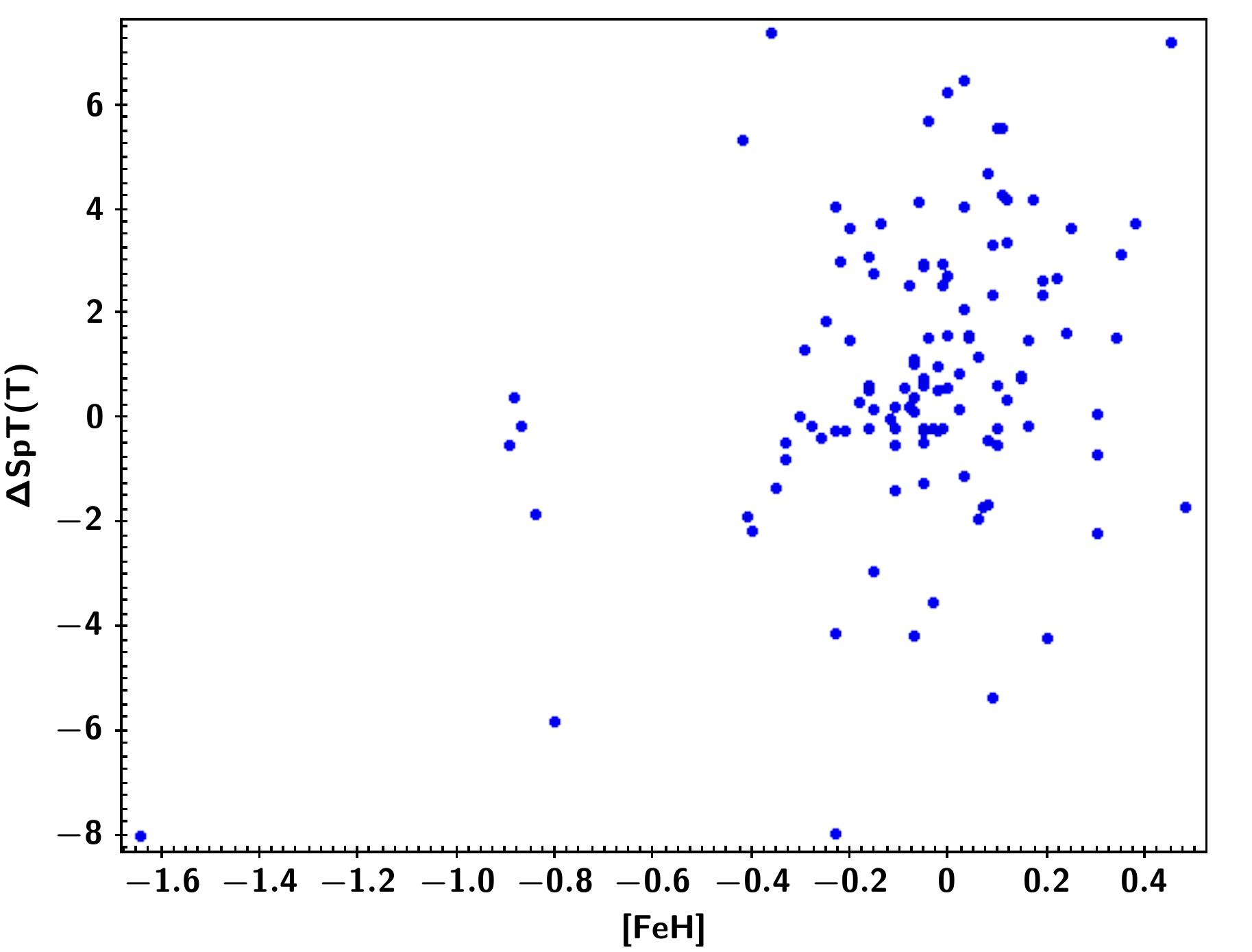}
\includegraphics[width=7cm]{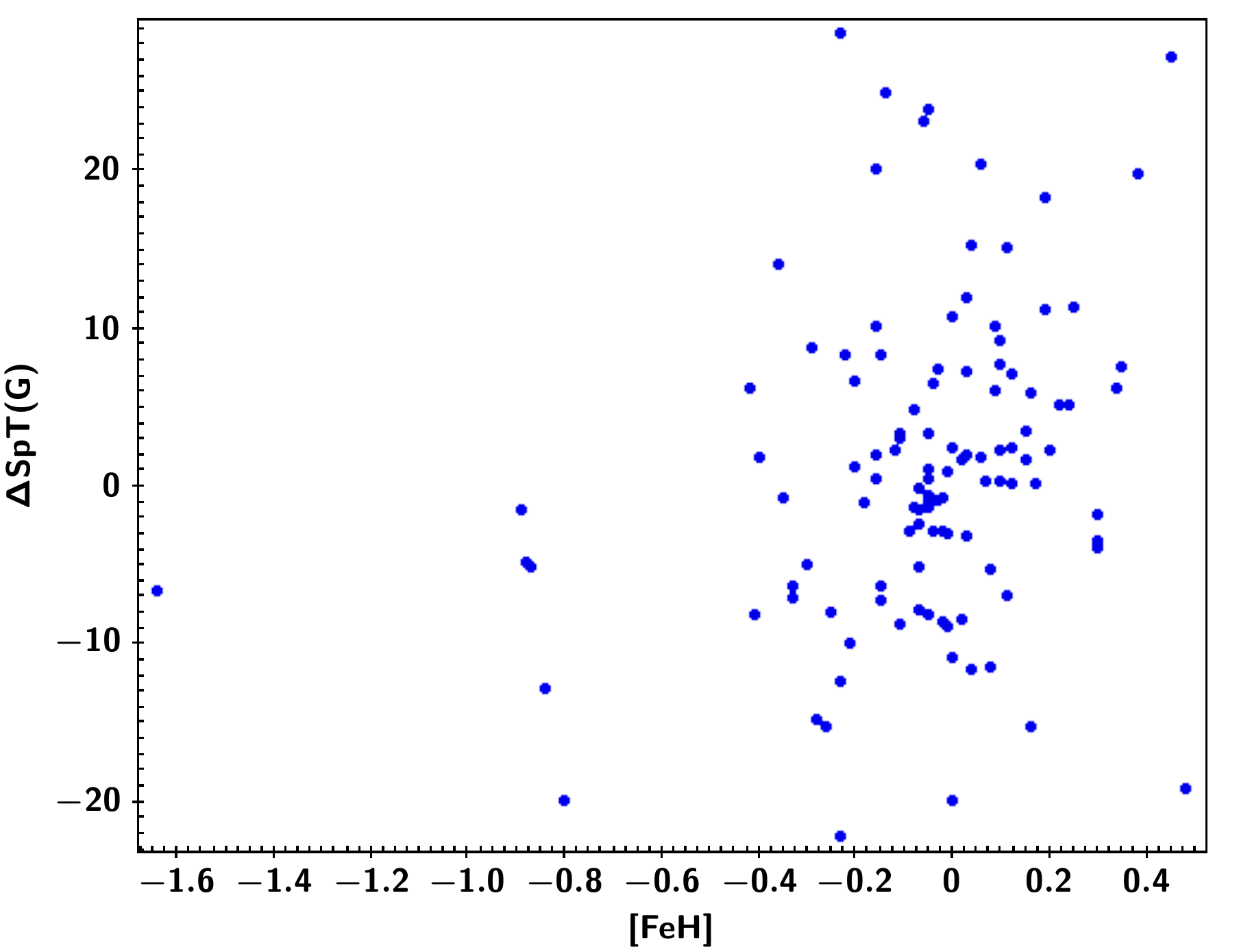}
\caption{Superiants. Difference between observed minus computed spectral class
vs. metallicity.
Left panel: $\Delta SpT(T)=SpT_{cat} - SpT(T_{\rm eff})$, where SpT$_{cat}$ is
the spectral class, catalogued in the empirical stellar spectral atlases,
and SpT($T_{\rm eff}$) is the spectral class, calculated from $T_{\rm eff}$ with Eq.~(T6).
Right panel: $\Delta SpT(G)=SpT_{cat} - SpT(\log g)$, where SpT($\log g$)
is the spectral class, calculated from $\log g$ with Eq.~(T8).
The Y-axis is graded so that one unit corresponds to one spectral sub-class
(e.g., the difference between A1 and A2).
}
\label{fig:s-feh-dS}
\end{figure}

To calculate these corrections,
we have selected giant and supergiant stars with known spectral class,
effective temperature $T_{\rm eff}$, surface gravity $\log g$, and metallicity [Fe/H]
from the empirical stellar spectral atlases
ELODIE~\citep{2007astro.ph..3658P},
Indo-US~\citep{2004ApJS..152..251V},
MILES~\citep{2011A&A...532A..95F},
and STELIB~\citep{2003A&A...402..433L}.
Besides, three giants from \citep{2010A&ARv..18...67T} were added.

After a preliminary analysis,
some stars were removed from the list.
The following giant stars were excluded from the consideration:

\begin{itemize}
\item HD~216131=$\mu$~Peg has $T_{\rm eff}$=4950~K, which is a little hotter
than what is required for M2III (ELODIE), however,
is perfectly consistent with G8III (Indo-US, MILES).

\item HD~130322 seems too dense ($\log g$=4.54) for its
spectral class K0III (ELODIE). According to SIMBAD, its spectral class is K0V.

\item Conversely, HD~37202 seems too rarefied ($\log g$=2.47) for its
spectral class B4IIIp (ELODIE). Indeed, according to
General Catalogue of Stellar Spectral Classifications~\citep{2014yCat....102023S}, it belongs
to the bright giant sequence and has spectral class B3IIp.

\item The same can be said for HD~190390.
According to SIMBAD, its spectral class is F2II, which is more consistent
with surface gravity value $\log g$=1.25 than F1III (Indo-US).

\item Besides, we have excluded HD~196777=$\upsilon$~Cap, a variable star,
which looks too hot ($T_{\rm eff}$=10500~K) and too dense ($\log g$=4.00)
for its spectral class M1III (Indo-US).
\end{itemize}

Also, we have excluded from further consideration some supergiants.

\begin{itemize}
\item $T_{\rm eff}$ of HD~217476 is 8320 K, which seems too large for
G0Iab (ELODIE).
The values listed in MILES look more self-consistent (G4Ia, $T_{\rm eff}$=5100~K).

\item Spectral class G4Ia of HD~6474 seems too late for its $T_{\rm eff}$=6240~K (MILES),
while General Catalogue of Stellar Spectral Classifications~\citep{2014yCat....102023S}
lists for this star spectral classes
from F8Ia to G5.

\item Spectral class K0Iab is catalogued for HD~104893, but, according to SIMBAD
its spectral class is F8/G2.
\end{itemize}

For the remaining 783 giants and 122 supergiants (their [Fe/H] distributions
are shown in Fig.~\ref{fig:feh}), we have computed
spectral classes from $T_{\rm eff}$ and $\log g$ with Eqs.~(T10) and~(T12) for giants,
and with Eqs.~(T6) and~(T8) for supergiants. The resulting values,
SpT($T_{\rm eff}$) and SpT($\log g$) were compared with spectral classes SpT$_{cat}$,
catalogued in the atlases. The difference O-C (observed minus computed) spectral class
is shown in Figs.~\ref{fig:g-feh-dS} and~\ref{fig:s-feh-dS} as a function of metallicity.
The Y-axis in Figs.~\ref{fig:g-feh-dS} and~\ref{fig:s-feh-dS}
is graded so that one unit corresponds to one spectral sub-class
(e.g., the difference between A1 and A2).

One can see that the difference between observed minus computed spectral class
correlates with metallicity. For giant stars, the spectral class value, computed
from $T_{\rm eff}$ with Eq.~(T10),
should be increased by the value of $\Delta SpT(T)$, where
\begin{equation}
\Delta SpT(T) \equiv SpT_{cat} - SpT(T_{\rm eff}) = 2.04 [Fe/H] - 0.34,
\label{equ:g-feh-dST}
\end{equation}
correlation coefficient is 0.44 (see Fig.~\ref{fig:g-feh-dS}, left panel).
Here SpT$_{cat}$ is the spectral class, catalogued in the empirical stellar spectral atlases,
and SpT($T_{\rm eff}$) is the spectral class, calculated from $T_{\rm eff}$ with Eq.~(T10).

Similarly to Eq.~(\ref{equ:g-feh-dST}), the spectral class value, computed
from $\log g$ with Eq.~(T12),
should be increased by the value of $\Delta SpT(G)$, where
\begin{equation}
\Delta SpT(G) \equiv SpT_{cat} - SpT(\log g) = 3.84 [Fe/H] + 0.73,
\label{equ:g-feh-dSG}
\end{equation}
correlation coefficient is 0.26 (see Fig.~\ref{fig:g-feh-dS}, right panel).
Here SpT$_{cat}$ is the spectral class, catalogued in the empirical stellar spectral atlases,
and SpT($\log g$) is the spectral class, calculated from $\log g$ with Eq.~(T12).
Note that the standard deviation of $\Delta SpT(G)$ (std.dev.=6.2)
is much larger than one of $\Delta SpT(T)$ (std.dev.=1.9).
Eqs.~(\ref{equ:g-feh-dST}) and~(\ref{equ:g-feh-dSG}) are valid for $-3 \le [Fe/H] \le 0.85$

The analogous relations for supergiants are not so obvious.
The following conclusions can be drawn from Fig.~\ref{fig:s-feh-dS} (left panel):
\begin{equation}
\Delta SpT(T) \equiv SpT_{cat} - SpT(T_{\rm eff}) =\begin{cases}
0,   &\text{for }-0.42 \le [Fe/H] \le 0.48;\\
-1.6, &\text{for }-0.9 \le [Fe/H] \le -0.8.
\end{cases}
\label{equ:s-feh-dST}
\end{equation}
Here SpT($T_{\rm eff}$) is the spectral class, calculated from $T_{\rm eff}$
with Eq.~(T6).

The lack of data does not allow us to draw definite conclusions beyond these ranges.
In particular, the data on the lowest metallicity stars is too scarce to make conclusions:
the only supergiant star with lower metallicity
in our sample, HD~103036 = TY~Vir ([Fe/H]=-1.64) is a long-period semiregular variable star.

The large standard deviation of $\Delta SpT(G)$ (std.dev.=9.8, see Fig.~\ref{fig:s-feh-dS}, right panel)
indicates that Eq.~(T8) should be applied with caution, and only in the range
$-0.42 \le [Fe/H] \le 0.48$ (where $\Delta SpT(G)$=0 can be used, as a first approximation).

\section{Verification of results with LAMOST data}
\label{sec:lamost}

\begin{figure}
\centering
\includegraphics[width=8cm]{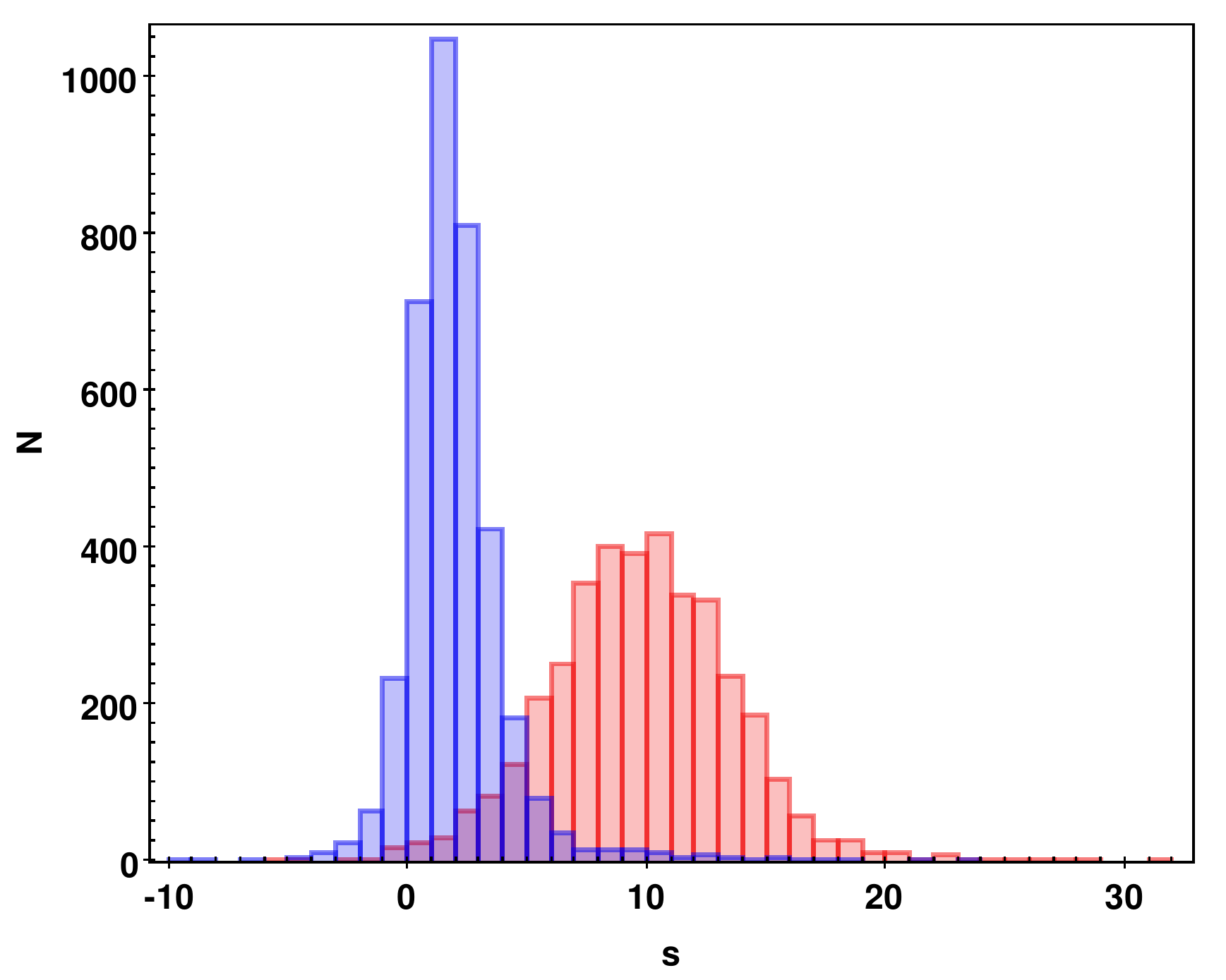}
\caption{Comparison of our results for LAMOST giant stars.
The red histogram is the $s_t$-$s_g$ distribution, and
the blue histogram is the $\bar s$ - $s_{\footnotesize LAMOST}$ distribution.
See Section~\ref{sec:lamost} for details.}
\label{fig:lamost}
\end{figure}

To verify our results, we have used data from LAMOST,
the largest source, containing independently determined spectral class
and atmospheric parameters for tens of thousands of stars.

The LAMOST (Large Sky Area Multi-Object Fiber Spectroscopic Telescope)
is a special reflecting Schmidt telescope
\citep{2012RAA....12.1197C},
which can observe 4000 spectra simultaneously in a
single exposure. Consequently, LAMOST has great potential to
efficiently survey a large volume of space for stars and galaxies
\citep{2012RAA....12..723Z}.

LAMOST AFGK stars catalogue \citep{2015RAA....15.1095L}
contains 5,843,107 objects (DR6 V1).
Atmospheric parameters and spectral class are
determined for them, and part of the objects have also luminosity classes.
LAMOST stellar spectral classification procedure is described by
\cite{2014AJ....147..101W}.

Main sequence, giant and subgiant stars are included in the LAMOST
classification scheme.
Supergiants are not indicated in the LAMOST catalogue.

\subsection{Giants}

Among 5,843,107 catalogued objects, 3716 stars are classified as
(mostly late-A) giants. For those stars
we have estimated spectral classes independently from
$T_{\rm eff}$ and $\log g$ values from the catalogue
(using Eqs.~(T10) and~(T12) from Table~\ref{tab:main}, respectively),
and compared them with catalogued spectral classes.

The result of our comparison for those 3716 stars is summarized in Fig.~\ref{fig:lamost}.
We have calculated a difference $s_t-s_g$, where
$s_t$ and $s_g$ represent
spectral class code estimated from effective temperature
(Table~\ref{tab:main}, Eq.~(T10))
and surface gravity
(Table~\ref{tab:main}, Eq.~(T12)), respectively.
Again, here
spectral class is coded as follows: 3 for O3, ..., 10 for B0, ..., 60 for M0.
We have also calculated an average value $\bar s = (s_t+s_g)/2$
and compared this value with catalogued spectral class $s_{\footnotesize LAMOST}$.
The red histogram in Fig.~\ref{fig:lamost} is the $s_t$-$s_g$ distribution, and
the blue histogram is the $\bar s$ - $s_{\footnotesize LAMOST}$ distribution.
One can see that whereas $s_t$ is consistently about one class later than $s_g$,
the average value $\bar s$, estimated from effective temperature and surface gravity,
in most cases differs by not more than three spectral sub-classes
from the catalogued spectral class $s_{\footnotesize LAMOST}$.
For 78\% and 90\% of stars that difference does not exceed three
and four spectral sub-classes, respectively
(see the blue histogram in Fig.~\ref{fig:lamost}).
Mean value for the difference $\bar s$-$s_{\footnotesize LAMOST}$ is 1.99.
It means that, e.g., for A7III star our procedure predicts, on average,
A9III spectral class.

\subsection{Main sequence stars}

\begin{figure}
\centering
\includegraphics[width=6.5cm]{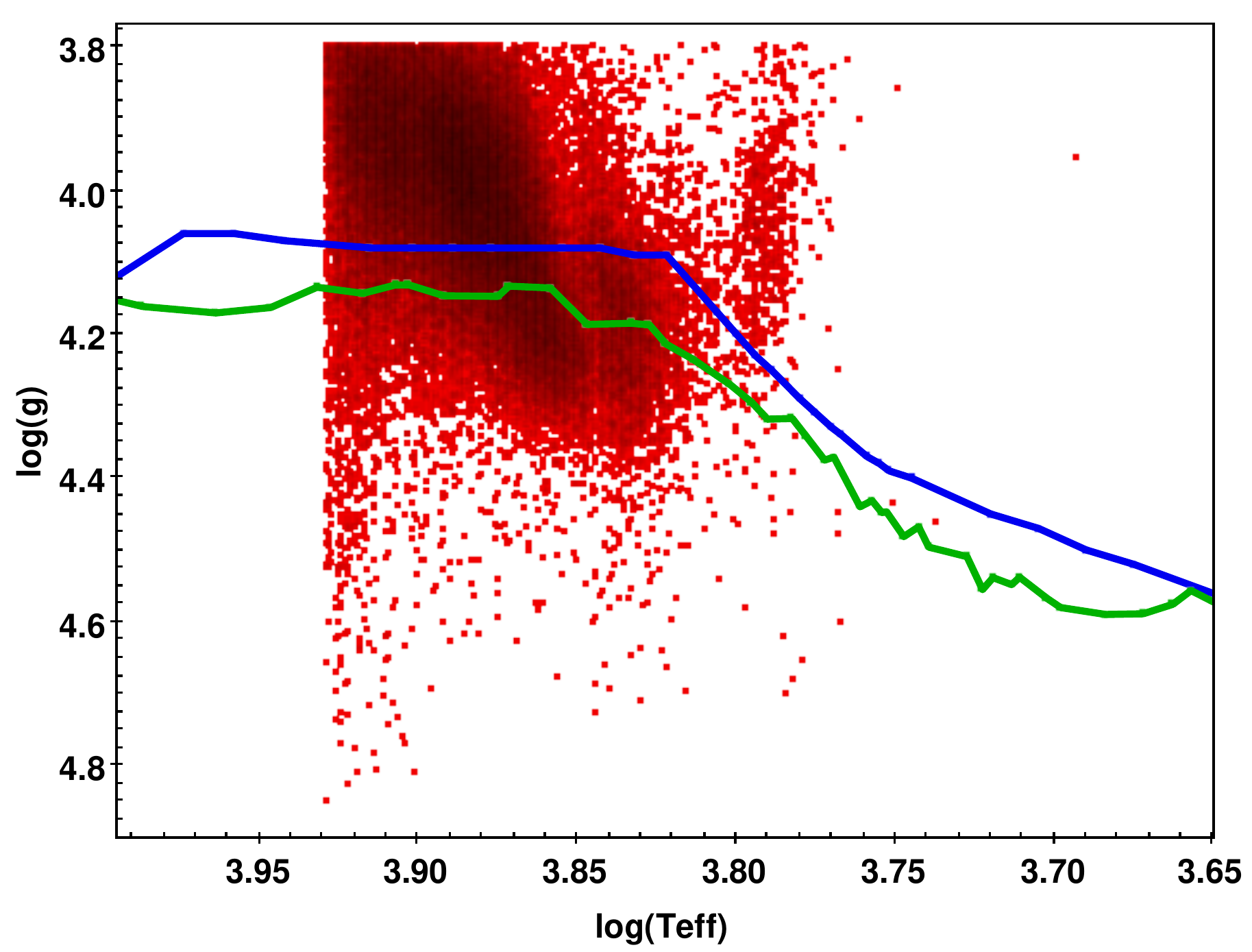}
\caption{LAMOST DR6 main sequence stars on the $\log T_{\rm eff}$ -- $\log g$ plot
(red dots). Blue and green curves represent corresponding relations
by \cite{2018MNRAS.479.5491E} and \cite{2013ApJS..208....9P}, respectively.
Note that the X- and Y-axes are flipped.}
\label{fig:lamostV}
\end{figure}

\begin{figure}
\centering
\includegraphics[width=7cm]{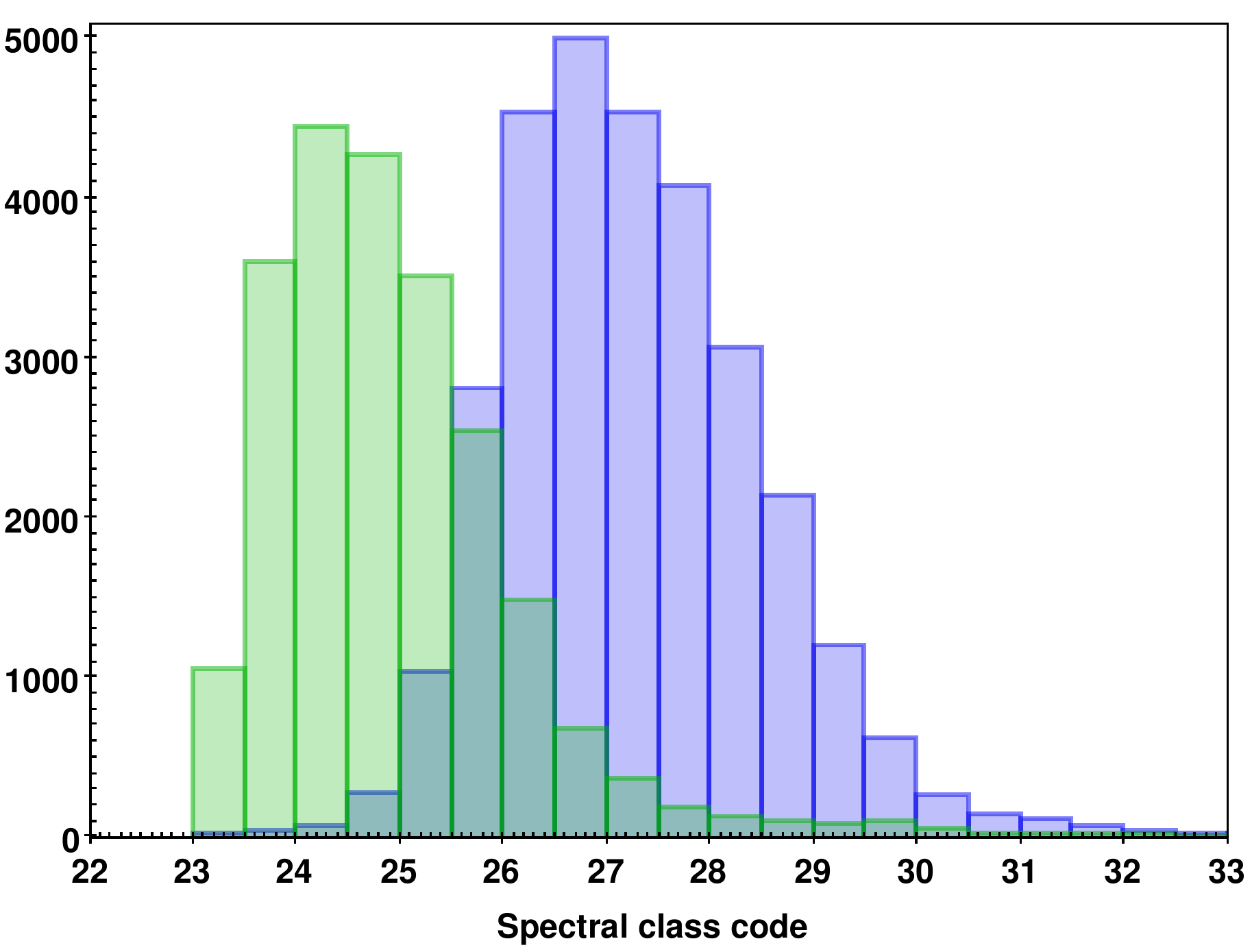}
\includegraphics[width=7cm]{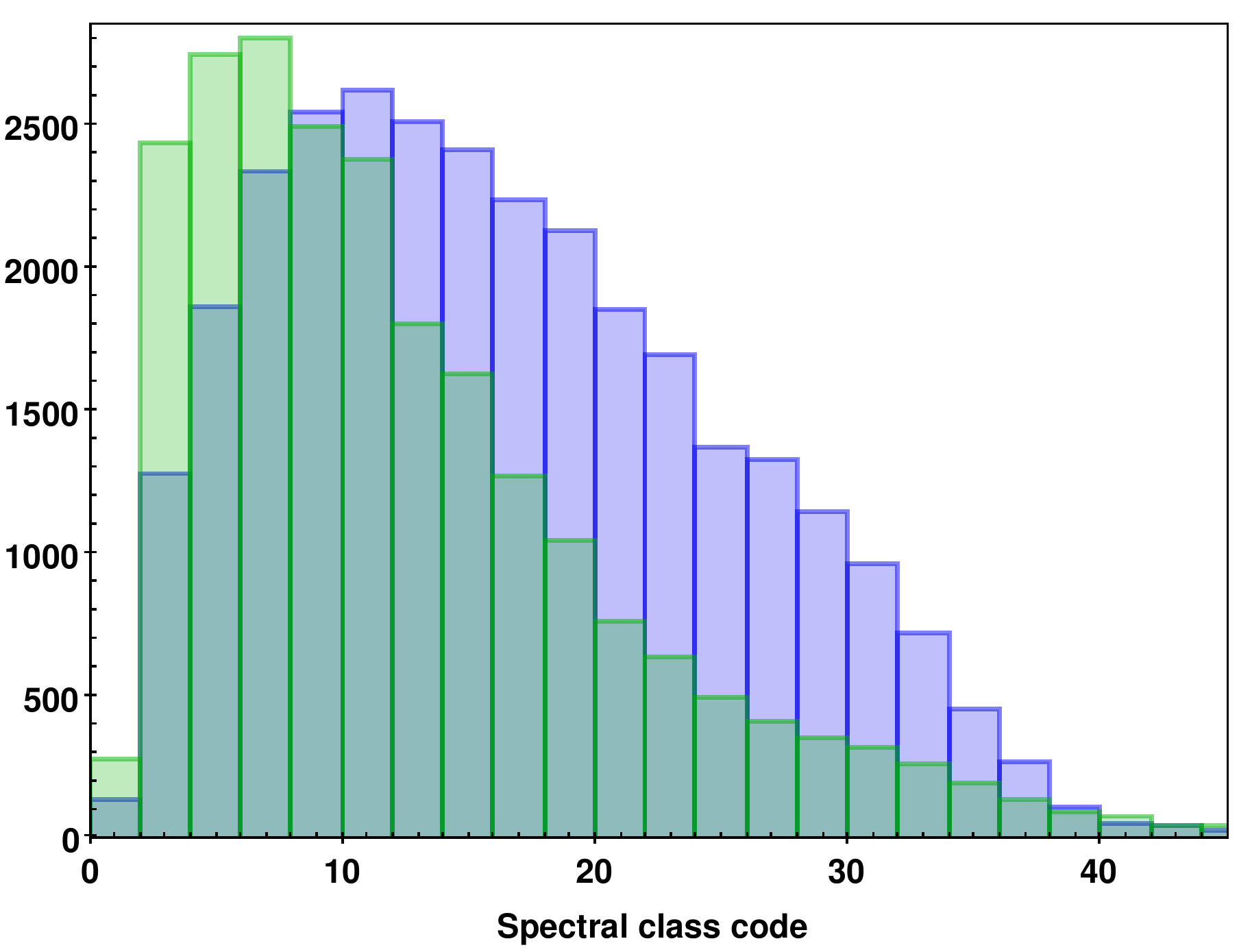}
\caption{Distribution of spectral class computed from $T_{\rm eff}$ (Eq.~(T2), left panel)
and $\log g$ (Eq.~(T4), right panel) for 22696 A5 stars (green bars) and 30102 A7 stars
(blue bars). The agreement of spectral classes computed from $T_{\rm eff}$ with
catalogued ones is perfect, while for the $\log g$ case,
the difference between the calculated and cataloged values is significant.
Spectral class is coded as follows: 3 for O3, ..., 20 for A0, ..., 40 for G0.}
\label{fig:A5A7}
\end{figure}

LAMOST DR6 AFGK catalogue contains 63,220 stars marked as having luminosity class V.
Part of them have $\log g<3.8$ (up to $\log g=0.784$), which is
beyond the applicability of Eq.~(T4) (see Table~\ref{tab:main}), and they
have been excluded from further analysis.
All of the remaining 58652 stars have spectral class A, and 90\%
of them belong to A5 or A7 sub-class.

The spread of these stars on the $\log T_{\rm eff}$ -- $\log g$ plot
is shown in Fig.~\ref{fig:lamostV}, together with corresponding relations
by \cite{2018MNRAS.479.5491E} and \cite{2013ApJS..208....9P}.
Note that according to \cite{2018MNRAS.479.5491E} and \cite{2013ApJS..208....9P} relations, A-stars range
is limited to approximately the following values
$4.00 \ge \log T_{\rm eff} \ge 3.87$ and
$4.0 \le \log g \le 4.2$.
It can be seen that some of the LAMOST stars are located far
(sometimes quite far) beyond this range.

For those 58652 main-sequence stars
we have estimated spectral classes independently from
$T_{\rm eff}$ and $\log g$ values from the catalogue
(using Eqs.~(T2) and~(T4) from Table~\ref{tab:main}, respectively),
and compared them with catalogued spectral classes.
The result of this comparison for A5 and A7 stars is summarized in Fig.~\ref{fig:A5A7}.

Spectral class in Fig.~\ref{fig:A5A7} is coded as follows: 3 for O3, ..., 20 for A0, ..., 40 for G0.
It can be seen that spectral classes, computed from $T_{\rm eff}$ (left panel), reproduce
the original data perfectly. The mean value for the histograms is
24.92 (std.dev.=1.28) for A5 stars, and 27.23 (std.dev.=1.31) for A7 stars.
At the same time, spectral classes, computed from $\log g$ (right panel),
are significantly different, on average, from the catalogued values.
The histograms predict much earlier spectral classes (on average, B3 for the LAMOST A5 stars,
and B7 for the LAMOST A7 stars), and demonstrate 
a much larger value spread (std.dev. are 8.75 and 8.94, respectively).
It can be suggested that LAMOST's spectral classes and $T_{\rm eff}$ values
are self-consistent for MS-stars, while $\log g$ values demonstrate disagreement, at least for
some stars.

\section{Conclusions}
\label{sec:conclusions}

We have approximated spectral class -- atmospheric parameters relations
for main sequence, giant and supergiant stars, using both observational data
and published calibration tables. To judge our results, we compare them
with previous representations. It appears that the agreement is satisfactory.
In some regions, our results deviate from the previous data.
This is a consequence of our inclusion of newer data that were not
available at the time the earlier interpolation tables were compiled.

We have verified the relations
for giants and MS-stars with LAMOST spectral data, and
the results of the comparison of our estimations with observations
had been quite satisfactory.
Here we consider our relations to be a formal, rough tool for spectral class
estimation, so we do not apply basic laws of statistics (such as the Cromwell rule),
i.e., we do not estimate the probability of the correctness of the decision.

The obtained results can be of use for estimation of effective temperature and
surface gravity from MK spectral class or for estimation of MK spectral class
from effective temperature, surface gravity and metallicity.
In particular, it could help to use spectral energy distributions from theoretical stellar
atlases for given ($T_{\rm eff}$, $\log g$, [Fe/H]) values,
and assign the resulting values to corresponding spectral classes.
That procedure is necessary, e.g., for estimation of absolute
stellar magnitudes in one or the other photometric system
used in modern sky surveys. It should be noted also that
the use of MK spectral classification provides us with an easy
(and less time-consuming) way to estimating the parameters of stars,
in contrast to the ($T_{\rm eff}$, $\log g$) -- model selection.

\begin{acknowledgements}
We are grateful to Eric Mamajek for helpful comments and
to our reviewer whose constructive comments greatly helped us to improve the paper.
OM thanks the CAS President's International Fellowship Initiative (PIFI).
This work has been partially supported by NSFC/RFBR grant 20-52-53009.
Guoshoujing Telescope (the Large Sky Area Multi-Object Fiber Spectroscopic Telescope LAMOST) is a National Major Scientific Project built by the Chinese Academy of Sciences. Funding for the project has been provided by the National Development and Reform Commission. LAMOST is operated and managed by the National Astronomical Observatories, Chinese Academy of Sciences.
This research has made use of 
the SIMBAD database, operated at CDS, Strasbourg, France, and
NASA's Astrophysics Data System.
This research made use of TOPCAT, an interactive graphical viewer and editor for tabular data \citep{2005ASPC..347...29T}
The acknowledgements were compiled using the Astronomy Acknowledgement Generator.
\end{acknowledgements}

\bibliographystyle{raa}
\bibliography{mksz}

\begin{thebibliography}{37}
\providecommand\natexlab[1]{#1}
\providecommand\JournalTitle[1]{#1}

\bibitem[{Allen}(1976)]{1976asqu.book.....A}
{Allen}, C.~W. 1976, {Astrophysical Quantities}

\bibitem[{Andersen}(1991)]{1991A&ARv...3...91A}
{Andersen}, J. 1991, \aapr, 3, 91

\bibitem[{Castelli} \& {Kurucz}(2003)]{2003IAUS..210P.A20C}
{Castelli}, F., \& {Kurucz}, R.~L. 2003, in IAU Symposium, Vol. 210, Modelling
  of Stellar Atmospheres, ed. N.~{Piskunov}, W.~W. {Weiss}, \& D.~F. {Gray},
  A20

\bibitem[{Covey} {et~al.}(2007)]{2007AJ....134.2398C}
{Covey}, K.~R., {Ivezi{\'c}}, {\v Z}., {Schlegel}, D., {et~al.} 2007, \aj, 134,
  2398

\bibitem[{Cui} {et~al.}(2012)]{2012RAA....12.1197C}
{Cui}, X.-Q., {Zhao}, Y.-H., {Chu}, Y.-Q., {et~al.} 2012, Research in Astronomy
  and Astrophysics, 12, 1197

\bibitem[{de Jager} \& {Nieuwenhuijzen}(1987)]{1987A&A...177..217D}
{de Jager}, C., \& {Nieuwenhuijzen}, H. 1987, \aap, 177, 217

\bibitem[{Eker} {et~al.}(2018)]{2018MNRAS.479.5491E}
{Eker}, Z., {Bak{\i}{\c s}}, V., {Bilir}, S., {et~al.} 2018, \mnras, 479, 5491

\bibitem[{Falc{\'o}n-Barroso} {et~al.}(2011)]{2011A&A...532A..95F}
{Falc{\'o}n-Barroso}, J., {S{\'a}nchez-Bl{\'a}zquez}, P., {Vazdekis}, A.,
  {et~al.} 2011, \aap, 532, A95

\bibitem[{Findeisen} \& {Hillenbrand}(2010)]{2010AJ....139.1338F}
{Findeisen}, K., \& {Hillenbrand}, L. 2010, \aj, 139, 1338

\bibitem[{Findeisen} {et~al.}(2011)]{2011AJ....142...23F}
{Findeisen}, K., {Hillenbrand}, L., \& {Soderblom}, D. 2011, \aj, 142, 23

\bibitem[{Gustafsson} {et~al.}(2008)]{2008A&A...486..951G}
{Gustafsson}, B., {Edvardsson}, B., {Eriksson}, K., {et~al.} 2008, \aap, 486,
  951

\bibitem[{Harmanec}(1988)]{1988BAICz..39..329H}
{Harmanec}, P. 1988, Bulletin of the Astronomical Institutes of Czechoslovakia,
  39, 329

\bibitem[{Hoffleit} \& {Jaschek}(1991)]{1991bsc..book.....H}
{Hoffleit}, D., \& {Jaschek}, C. 1991, {The Bright star catalogue}

\bibitem[{Johnson}(1966)]{1966ARA&A...4..193J}
{Johnson}, H.~L. 1966, \araa, 4, 193

\bibitem[{Kim} \& {Moon}(2011)]{2011AJ....141..118K}
{Kim}, C., \& {Moon}, B.-K. 2011, \aj, 141, 118

\bibitem[{Kim} \& {Moon}(2014)]{2014Ap&SS.351..229K}
{Kim}, C., \& {Moon}, B.-K. 2014, \apss, 351, 229

\bibitem[{Kraus} \& {Hillenbrand}(2007)]{2007AJ....134.2340K}
{Kraus}, A.~L., \& {Hillenbrand}, L.~A. 2007, \aj, 134, 2340

\bibitem[{Le Borgne} {et~al.}(2003)]{2003A&A...402..433L}
{Le Borgne}, J.-F., {Bruzual}, G., {Pell{\'o}}, R., {et~al.} 2003, \aap, 402,
  433

\bibitem[{Lejeune} {et~al.}(1997)]{1997A&AS..125..229L}
{Lejeune}, T., {Cuisinier}, F., \& {Buser}, R. 1997, \aaps, 125, 229

\bibitem[{Luo} {et~al.}(2015)]{2015RAA....15.1095L}
{Luo}, A.-L., {Zhao}, Y.-H., {Zhao}, G., {et~al.} 2015, Research in Astronomy
  and Astrophysics, 15, 1095

\bibitem[{Malkov} {et~al.}(2010)]{2010MNRAS.401..695M}
{Malkov}, O.~Y., {Sichevskij}, S.~G., \& {Kovaleva}, D.~A. 2010, \mnras, 401,
  695

\bibitem[{Malkov} {et~al.}(2018{\natexlab{a}})]{2018OAst...27...62M}
{Malkov}, O., {Karpov}, S., {Kilpio}, E., {et~al.} 2018{\natexlab{a}}, Open
  Astronomy, 27, 62

\bibitem[{Malkov} {et~al.}(2018{\natexlab{b}})]{2018Galax...7....7M}
{Malkov}, O., {Karpov}, S., {Kovaleva}, D., {et~al.} 2018{\natexlab{b}},
  Galaxies, 7, 7

\bibitem[{Pecaut} \& {Mamajek}(2013)]{2013ApJS..208....9P}
{Pecaut}, M.~J., \& {Mamajek}, E.~E. 2013, \apjs, 208, 9

\bibitem[{Pecaut} {et~al.}(2012)]{2012ApJ...746..154P}
{Pecaut}, M.~J., {Mamajek}, E.~E., \& {Bubar}, E.~J. 2012, \apj, 746, 154

\bibitem[{Popper}(1980)]{1980ARA&A..18..115P}
{Popper}, D.~M. 1980, \araa, 18, 115

\bibitem[{Prugniel} {et~al.}(2007)]{2007astro.ph..3658P}
{Prugniel}, P., {Soubiran}, C., {Koleva}, M., \& {Le Borgne}, D. 2007,
  astro-ph/0703658

\bibitem[{Sichevskij} {et~al.}(2014)]{2014AstBu..69..160S}
{Sichevskij}, S.~G., {Mironov}, A.~V., \& {Malkov}, O.~Y. 2014, Astrophysical
  Bulletin, 69, 160

\bibitem[{Sichevskiy} {et~al.}(2013)]{2013AN....334..832S}
{Sichevskiy}, S.~G., {Mironov}, A.~V., \& {Malkov}, O.~Y. 2013, Astronomische
  Nachrichten, 334, 832

\bibitem[{Skiff}(2014)]{2014yCat....102023S}
{Skiff}, B.~A. 2014, VizieR Online Data Catalog, B/mk

\bibitem[{Smalley} \& {Dworetsky}(1995)]{1995A&A...293..446S}
{Smalley}, B., \& {Dworetsky}, M.~M. 1995, \aap, 293, 446

\bibitem[{Strai{\v z}ys}(1992)]{1992msp..book.....S}
{Strai{\v z}ys}, V. 1992, {Multicolor stellar photometry}

\bibitem[{Taylor}(2005)]{2005ASPC..347...29T}
{Taylor}, M.~B. 2005, in Astronomical Society of the Pacific Conference Series,
  Vol. 347, Astronomical Data Analysis Software and Systems XIV, ed.
  P.~{Shopbell}, M.~{Britton}, \& R.~{Ebert}, 29

\bibitem[{Torres} {et~al.}(2010)]{2010A&ARv..18...67T}
{Torres}, G., {Andersen}, J., \& {Gim{\'e}nez}, A. 2010, \aapr, 18, 67

\bibitem[{Valdes} {et~al.}(2004)]{2004ApJS..152..251V}
{Valdes}, F., {Gupta}, R., {Rose}, J.~A., {Singh}, H.~P., \& {Bell}, D.~J.
  2004, \apjs, 152, 251

\bibitem[{Wei} {et~al.}(2014)]{2014AJ....147..101W}
{Wei}, P., {Luo}, A., {Li}, Y., {et~al.} 2014, \aj, 147, 101

\bibitem[{Zhao} {et~al.}(2012)]{2012RAA....12..723Z}
{Zhao}, G., {Zhao}, Y.-H., {Chu}, Y.-Q., {Jing}, Y.-P., \& {Deng}, L.-C. 2012,
  Research in Astronomy and Astrophysics, 12, 723

\end{thebibliography}

\label{lastpage}
\end{document}